# MEASUREMENT AND MANIPULATION OF BETA FUNCTIONS IN THE FERMILAB BOOSTER

M. McAteer, S. Kopp, The University of Texas at Austin, Austin, TX 78712
E. Prebys, FNAL, Batavia, IL 60510

*Abstract*

In order to meet the needs of Fermilab's planned post-collider experimental program, the total proton throughput of the 8 GeV Booster accelerator must be nearly doubled within the next two years [1]. A system of 48 ramped corrector magnets has recently been installed in the Booster to help improve efficiency and allow for higher beam intensity without exceeding safe radiation levels [2].

We present the preliminary results of beta function measurements made using these corrector magnets. Our goal is to use the correctors to reduce irregularities in the beta function, and ultimately to introduce localized beta bumps to reduce beam loss or direct losses towards collimators.

## INTRODUCTION

The Booster's lattice is composed of 24 periods of combined function magnets in a FOFDOOD arrangement (Fig. 1). 48 new corrector magnet packages have recently been installed in the Booster, one in each short and long drift section. Each package contains a twelve-pole laminated core which has been wound to produce six independently controllable multipole elements: horizontal and vertical dipole, normal and skew quadrupole, and normal and skew sextupole. They are strong enough to move the beam by one centimeter and change the tunes by ±0.1 at 8 GeV.

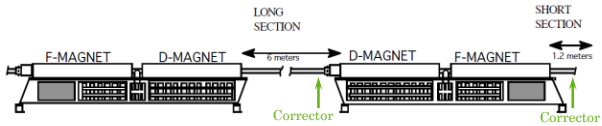

Figure 1. Arrangement of combined-function magnets in one period of the Booster.

We used the corrector magnets to measure the beta functions around the ring by introducing quadrupole errors and measuring the resulting tune shifts. For an uncoupled machine, a single quadrupole error $\delta q$ will cause an increase to the tune in one plane and a decrease in the other: $\delta \nu_x = \beta_x \cdot \delta q / 4\pi$, and $\delta \nu_y = -\beta_y \cdot \delta q / 4\pi$. The Booster, however, has significant transverse coupling, and in a coupled system the observable eigentunes are a superposition of the unperturbed x and y tunes. Near a $\nu_x - \nu_y \approx 0$ resonance the observable eigentunes are $\nu_\pm = (\nu_x + \nu_y \pm \sqrt{\kappa^2 + \Delta^2})/2$, where $\Delta = \nu_x - \nu_y$ is the separation between the uncoupled tunes and $\kappa$ describes the strength of the transverse coupling [3]. By setting $\nu_x \to \nu_x + \beta_x \cdot \delta q / 4\pi$ and $\nu_y \to \nu_y - \beta_y \cdot \delta q / 4\pi$, and then expanding to first order in $\kappa$, we obtain an expression for the tune shifts in terms of the beta functions, the coupling strength $\kappa$, and the quadrupole error $\delta q$:

$$\delta \nu_\pm = \frac{1}{8\pi}\left[\beta_x\left(1 \pm \frac{\Delta}{\sqrt{\kappa^2 + \Delta^2}}\right) + \beta_y\left(1 \mp \frac{\Delta}{\sqrt{\kappa^2 + \Delta^2}}\right)\right]\delta q \quad (1)$$

## METHOD

### Tune Measurement

Tune changes are the observable quantity measured to determine both coupling strength and beta functions. The tune in the Booster was found by estimating the peak of the continuous Fourier transform of the phased sum of 24 BPM readings around the ring [4]:

$$X(\nu) = \frac{1}{N}\sum_{n=1}^{N} e^{-2\pi i \nu (n-1)} x_n \quad (2)$$

The tune in the Booster changes rapidly throughout the acceleration cycle, so only N=64 turns were used in the tune measurements.

### Coupling Measurement

The measurement of the beta function requires knowledge of the coupling strength $\kappa$, which corresponds to the minimum possible eigentune separation, so we measured the tune separation throughout the early part of the acceleration cycle. We changed the horizontal tune by small amounts for a few milliseconds at a time, leaving the vertical tune unchanged, and measured the resulting eigentunes (Fig. 2). We fit the eigentune difference as a function of the horizontal tune shift to the hyperbola $\Delta \nu_\pm = \sqrt{\kappa^2 + \Delta^2}$ to find the minimum tune separation $\kappa$ (Fig. 3). We measured $\kappa$ at 500-turn increments through the first 7000 turns of the Booster cycle (Fig. 4).

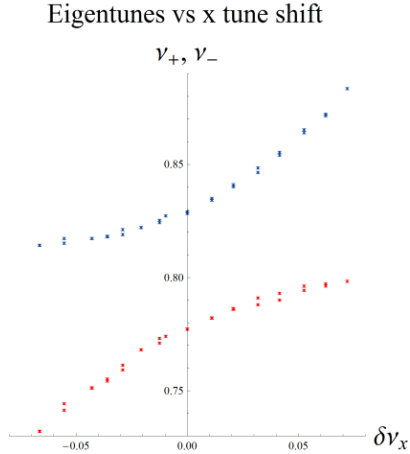

Figure 2. Eigentunes, measured as small changes were made to the unperturbed horizontal tune.

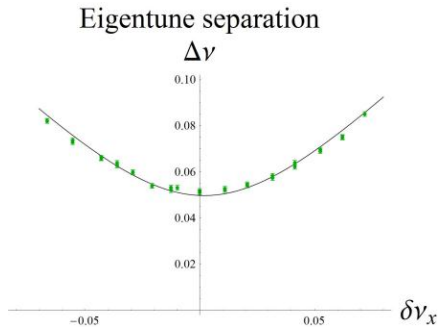

Figure 3. Separation between the eigentunes, with a hyperbolic fit to find the minimum tune separation κ.

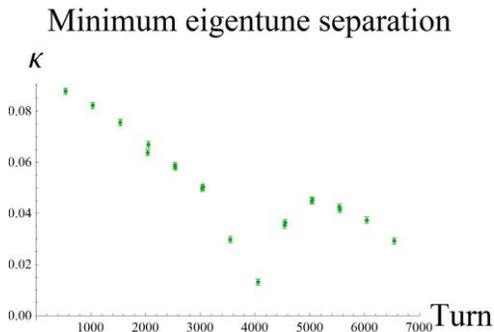

Figure 4. Measured minimum tune separation through the beginning of the Booster's acceleration cycle.

*Beta Function Measurements*

The beta functions were measured at the location of each corrector magnet by increasing the quad's current by 15 amps for a few milliseconds and measuring the resulting changes to the tunes. Small random variations in tune on each pulse are significant on the scale of the tune shifts that we are looking for, so it is necessary to take multiple measurements to reduce the effect of random tune variation. The initial, unmodified tune was taken to be the average tune measured over 20 pulses, and the tune was measured three times while each quad bump was in place in order to determine the average value for the tune shift caused by each quad error.

Figure 5 shows the tune shifts due to quadrupole errors in the short drift sections of the Booster. Because of the strong transverse coupling, a quad error causes an increase in both the x and y tunes, rather than an increase in the x tune and a decrease in the y tune as would be expected in an uncoupled machine.

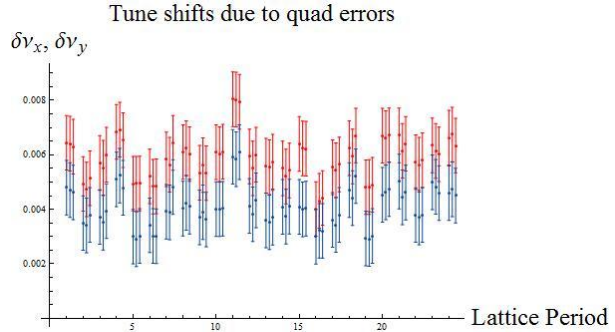

Figure 5. The x and y tune shifts caused by a single quad error in each short drift section of the Booster. Both tunes increase because coupling is strong.

The horizontal and vertical beta functions at the location of each corrector magnet were calculated from the measured tune shifts and the measured coupling strength using Eq. 1. The results are shown in figures 6 and 7.

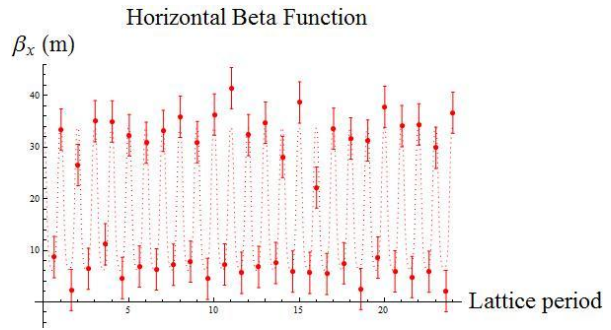

Figure 6. Measured horizontal beta functions before correction, shown with the design values.

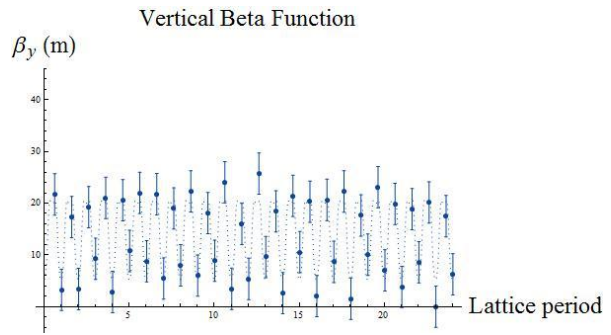

Figure 7. Measured vertical beta functions before correction, shown with the design values.

*Correction of Beta Functions*

We will attempt to correct the irregularities in the beta functions by adjusting each quad magnet to compensate for the measured beta deviation at each location. We use only the high-beta-plane correctors in each section; the 24 short-section magnets will be used to correct the horizontal beta function in the short sections, and the 24 long-section magnets will be used to correct the vertical beta function in the long sections.

We used a first-order approximation of the effects of all 24 small quad errors on the beta function to compute the needed corrections:

$$\Delta\vec{\beta} = \overline{M} \cdot \Delta\vec{q},$$
$$M_{ij} = \frac{\beta(s_i)\cdot\beta(s_j)}{2 Sin(2\pi\nu)} Cos\big(2|\psi(s_i)-\psi(s_j)|-2\pi\nu\big) \quad (3)$$

The correction currents calculated using this method are large enough that the first-order approximation of their effects on the beta function is only accurate to within ~95%, and the uncertainty is comparable to the variation (Fig. 8). We plan to repeat the process using many more tune shift measurements to reduce uncertainty in the calculated beta function values, and we expect to find that smaller corrections will be needed when the beta functions are measured more accurately. If the corrections calculated from the improved beta measurements are still very large, we will implement a scaled-down partial correction and attempt to even out the beta function through an iterative process.

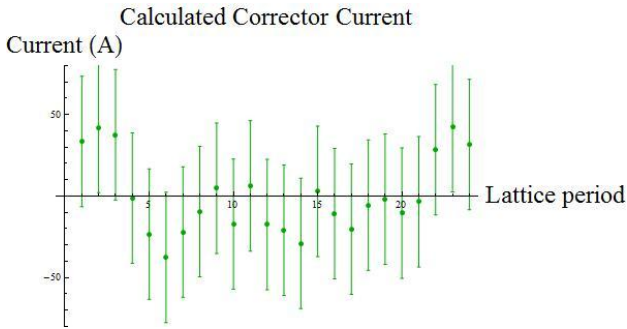

Figure 8. Long-section corrector quadrupole currents needed to compensate for the measured deviation of the vertical beta function.

## CONCLUSIONS

This method for measuring and correcting the beta functions may be useful, but the measurements are time-consuming; the process must be modified to allow us to take larger volumes of data to get a precise average measurement of tune shifts. We are working on automating the process to make the measurements more precise, and possibly to make beta function measurement a diagnostic tool readily available for operational use.

Since the variation in the beta functions appears to be no greater than about 20%, it is possible that no significant reductions to losses will be attained by correcting the beta functions. However, we also plan to use similar methods to make localized beta bumps, which should help to reduce losses around areas with restricted apertures or to steer losses towards collimators.